\documentclass[12pt,preprint]{aastex}

\begin{document}

\title{Quiescence and late time activity in collapsars due to critical angular momentum distributions}

\author{Diego Lopez-Camara\altaffilmark{1}}\altaffiltext{1}{Instituto de Ciencias
  Nucleares, UNAM, Apdo. Postal 70-543, M\'{e}xico D.F. 04510,
  MEXICO}

\begin{abstract}
Even thought a large amount of long gamma ray bursts (LGRBs) present quiescent periods, their origin remains unclear. 
In this talk, it is shown how different angular momentum distributions, as a function of the stellar radius), can lead to neutrino luminosity variability and the possibility of quiescent epochs in LGRBs.
\end{abstract}

\section{Introduction}

Collapsing massive stars produce Supernovae (SN) and are linked to the
production of Long Gamma Ray Bursts (LGRBs), providing clues to the progenitors and their environments. A key ingredient is the stellar rotation rate, impacting the energy release and outcome of the event following the implosion of the iron core. Stellar evolution considerations have shown that it is
non-trivial to have rapid rotation, as mass loss and magnetic fields
conspire to reduce the pre-SN angular momentum. 

Many studies of neutrino cooled accretion relevant for collapsars have
considered specific rotation laws that guarantee by a large margin the
formation of a centrifugal disk, because the angular velocity is
assumed to be nearly Keplerian, or because the absolute value of the
angular momentum given implies a circularization radius much larger
than the radius of the innermost stable circular orbit. However, the distributions of specific angular momentum considered were constant in the equatorial plane. This is unrealistic, as the specific angular momentum increases outwards in the core and envelope, with marked transitions at the boundaries between different
shells.

Here we explore how the distribution of angular momentum as a function
of radius can affect the qualitative properties of the accretion flow,
and hence the neutrino luminosity, accretion rate and energy
release. We pay particular attention to the form and rate of change of
rotation in the star with radius, and show that state transitions may
in principle produce observable consequences in LGRBs relevant to
variability and quiescent periods.

Woosley \& Heger 's (2006) 1D pre-SN models were taken as the initial conditions. The correspondent distributions were mapped to two dimensions assuming spherical symmetry, and the iron core was condensed onto a point mass at the origin representing a BH, producing a pseudo-Newtonian potential. The evolution is subsequently followed with the same numerical code used in Lopez-Camara et al (2009). 

\section{State transitions.}\label{sec:states}

The general trend in the distribution of specific angular momentum in pre-SN
models is for a rise through the core and envelope. We thus initially considered
angular momentum which increased linearly as a function of the stellar radii ($J(R)$). When $J(R)$ increased very slowly the result was a quasi-radial inflow (QRI); on the other hand, for a rapidly increasing $J(R)$ an accretion disk around the BH was produced.
Interestingly, cases which increased linearly -but in neither of the two previous regimes-, this is: an intermediate linearly increasing case, allowed momentary
appearance of a torus, which was accreted after a delay of $\simeq
0.1$~s by th BH. 

\begin{figure}[h!]
  \begin{center}
\includegraphics[width=0.4\textwidth]{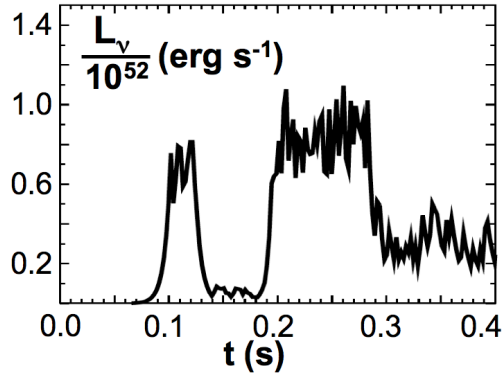}
\caption{Neutrino luminosity for the case where $J(R)$ has a constant background with two superimposed spikes.}
\label{fig:spikes} 
  \end{center}
\end{figure}

Since neighboring shells in pre-SN cores exhibit strong jumps in the
$J(R)$ superimposed on an increasing function of radius. To explore how this feature affects the flow properties in the collapsing star, we considered a constant background distribution just below the critical value to produce the accretion disk around the BH ($J_{\rm crit}$), and two narrow spikes with $J(R)$ well above $J_{\rm crit}$. The resulting neutrino luminosity is shown in
Figure~\ref{fig:spikes}. It is clear that multiple spikes in the distribution of specific angular momentum lead to clear transitions between the ``quasi-radial" low-$L_{\nu}$ and ``disk" high-$L_{\nu}$ state, with durations and delays correlated to the form and normalization of $J(R)$. 

\begin{figure}[h!]
  \begin{center}
\includegraphics[width=0.7\textwidth]{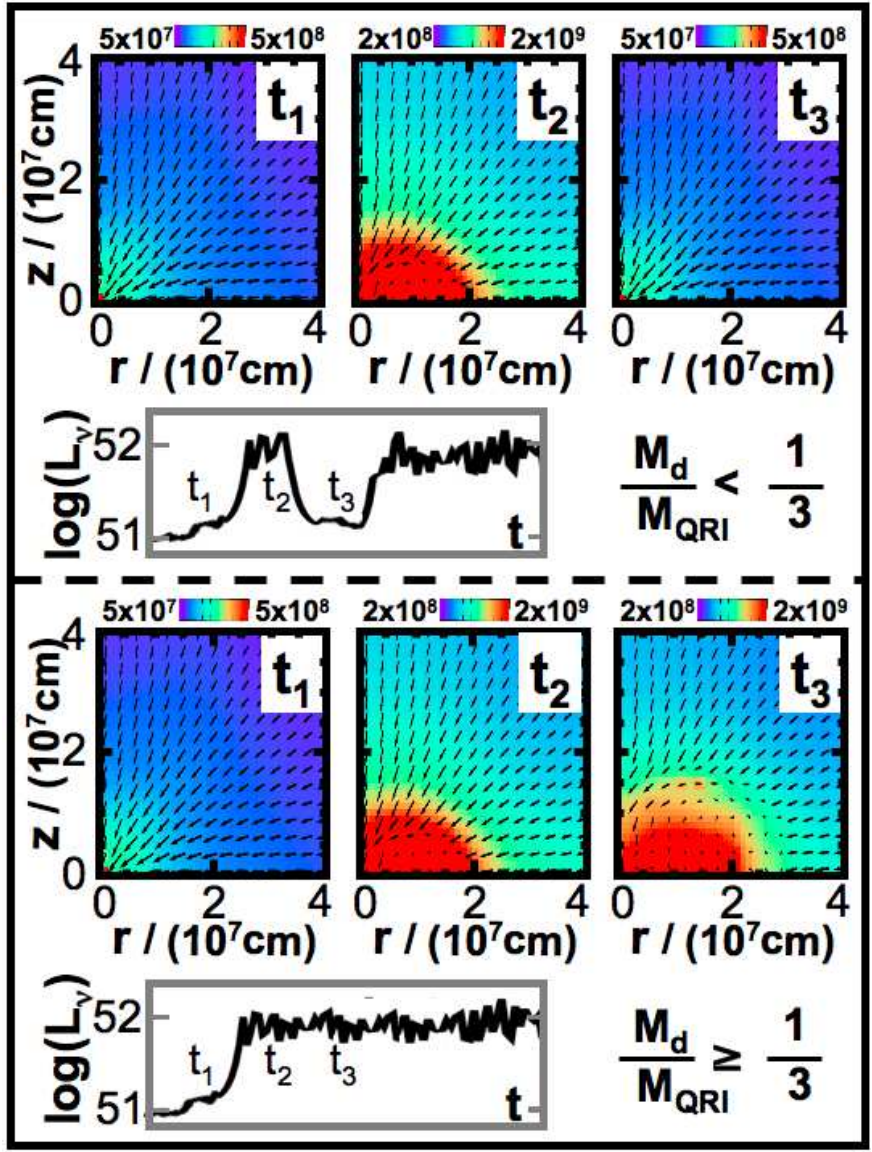}
\caption{The density and velocity field for the different scenarios.}
\label{fig:pathways} 
  \end{center}
\end{figure}

For the cases when the newly formed accretion disk had more than a third of the mass inside the QRI envelope which is falling onto it (case referred as: $\mu \geq 1/3$), then the initial disk absorbs the impact of
the infalling shell and survives. Thus, when a second spike would  approach the still existing disk it would simply add to the preexisting activity, leaving no place to quiescent periods. This is illustrated in the bottom panel of Figure~\ref{fig:pathways}. On the other hand, for cases when the disk has less than a third of that which is present inside the QRI envelope (case referred as: $\mu \leq 1/3$), then the accretion disk would be destroyed within a dynamical time scale. With this, when a second spike would reach the centrifugal barrier, a new disk would be created, persisting as long as the inflow has sufficient rotation (upper panel in Figure~\ref{fig:pathways}), and a quiescent epoch would be present.

The correspondent time scales can be estimated as the correspondent free fall time for each shell, thus for the case when there is a quiescent period, the initial active period lasts approximately between 1 and 3 second, while the quiescent period lasts from 1 to 15 seconds (depending on the preSN characteristics), prior to the main burst. The corresponding neutrino luminosities would be: $L_{\nu} \simeq 10^{51}$~erg~s$^{-1}$ for the quiescent period, and $L_{\nu} \simeq 10^{51}$~erg~s$^{-1}$ for the active periods.

\section{Discussion}\label{sec:discussion}

We wish to stress that while we have presented the neutrino luminosity
as a measure of energy output, it is by no means the only one
possible, and should be viewed here as a proxy for central engine
activity, like the mass accretion rate (with which it is closely
correlated). One could equally use $\dot{M}$ or the power output
through magnetic fields as a measure of the ability to drive
relativistic outflows. Our numerical scheme is geared towards
appropriate handling of thermodynamics and the neutrino emission, so
it is natural to rely on these properties when making quantitative
statements.

We note that just as not all LGRBs exhibit this behavior, clearly not
all progenitors are capable of producing such state transitions. Whether this can power precursor activity is another matter, requiring the initial episode of accretion to create a low density polar funnel in the star, which remains to be studied in the near future.

\end{document}